\documentclass[prl,preprint,showpacs]{revtex4-1}

\usepackage{mathrsfs}
\usepackage{graphics}%
\usepackage{dcolumn}%
\usepackage{bm}%
\usepackage{epsfig}%
\usepackage{tabularx}%
\usepackage{multirow}%
\usepackage{rotating}%
\usepackage{rotating}%
\usepackage[usenames]{color}

\begin{document}
\title{A generalized mean field theory of coarse-graining.}
\addtocounter{page}{0}

%\center{\bf{}}

\vskip0.2cm

\author{Luca Larini$^{*}$ and Vinod Krishna\footnotetext[1]{These authors have made equal contributions.}}\email{llarini@chem.ucsb.edu; krishna@ias.edu}
\affiliation{University of Utah, Department of Chemistry, UT 84112}

%\vskip0.2cm
\begin{abstract}
A general mean field theory is presented for the construction of equilibrium coarse grained 
models. Inverse methods that reconstruct microscopic models from low resolution experimental 
data can be derived as particular implementations of this theory. The theory also applies to 
the opposite problem of reduction, where relevant information is extracted from available 
equilibrium ensemble data. These problems are central to the construction of coarse grained 
representations of complex systems, and commonly used coarse graining methods are derived as 
particular cases of the general theory. 
\end{abstract}
\maketitle
%---------------------------------------
%            INTRODUCTION          
%---------------------------------------
Coarse-graining(CG) methods are widely used in both simulations and theory
to construct simplified models that render complex problems analytically 
and computationally tractable. Successful examples are the theory of
polymers \cite{doiEdwards}, the theory of liquid crystals \cite{deGennes} 
and hydrodynamics \cite{lamb}. In all these models the system is not described 
with an atomistic resolution, but only a few relevant properties of the system 
are retained to account for the essential physics required to describe the system.

Two general strategies are commonly used in the construction of CG models. The 
first involves finding a solution to an inverse problem, where some
macroscopic quantity is known and a microscopic structure of the
system is required that is consistent with this quantity. An example of this strategy is 
the reverse montecarlo (RMC) \cite{Lyub,muller} method where a radial distribution function (RDF) 
is provided as the target property that needs to be reproduced. A second class of methods
involve reducing the number of degrees of freedom present in the system. An
example of such an approach is the multiscale coarse-graining (MS-CG)
method\cite{Voth1}, where a CG model is built from an atomistic trajectory
generated from a molecular dynamics (MD) simulation. 
Inverse problems are of primary importance for the interpretation of experimental data, where 
macroscopic quantities are measured and a microscopic description of 
the system is required.\cite{Lyub,gura} On the other hand, an important challenge in data analysis 
is to develop systematic methods by which the reduction and organization of high resolution
data can be performed to extract the relevant pieces of information from available data.\cite{amadei,best}
Extracting relevant quantities is also essential for constructing simplified theoretical models that retain the
essential aspects of a complex physical process while discarding non-relevant degrees of freedom. 
This is particularly true for complex systems, such as biomolecules, where CG models help in 
exploring large scale properties.\cite{vbook} 

%+++++++++++++++++++++++++++++++++++++++
%            END OF INTRODUCTION         
%+++++++++++++++++++++++++++++++++++++++

Several techniques are available in the published literature that deal with the
above described problems, but a general formulation that can be used to
construct such methods does not currently exist. This work
establishes a general statistical mechanical formulation that can be
used to build such models, and which also serves as a general statistical 
mechanical framework within which the physics of such modeling strategies
can be interpreted and understood. The theory presented in this work offers a simple 
and general framework for a diverse class of methods which deal with the type of 
problems described above and is intimately connected with traditional classical mean 
field theories. Thus, the results from this work enable the application of techniques 
from statistical mechanics to develop novel and powerful methods for coarse-graining 
and the study of inverse problems in equilibrium statistical mechanics. 

\par A primary goal in the construction of a CG model is to provide a microscopic
description of the system under examination which can reproduce the ensemble 
average of a given set of macroscopic observables, $\langle\{\sigma_{i}\}\rangle$. 
An average of such a variable in the Gibbs-Boltzmann canonical ensemble can be computed as:
\begin{equation}
\label{average}
\langle\sigma_{i}\rangle=\frac{\int_D \mathrm{d}{\vec{R}}^{N}
  \mathrm{d}{\vec{P}}^{\,N}\ \sigma_{i}\ e^{-\beta
    \mathcal{H}(\vec{R}^{\,N},\vec{P}^{N})}}{\int_D \mathrm{d}
  \vec{R}^{N} \mathrm{d}{\vec{P}}^{N}\ e^{-\beta
    \mathcal{H}(\vec{R}^{N}, \vec{P}^{N})}}
\end{equation}
where $\beta=1/k_BT$, $k_B$ is the Boltzmann's constant and $T$ the
temperature of the system. $N$ is the number of particles in the system, ${\vec{R}}^{N}$
and ${\vec{P}}^{N}$ their coordinates and momenta respectively,
and  $\mathcal{H}$ the Hamiltonian describing the system. Thus the aim of a CG
methodology is to build an effective Hamiltonian $\mathcal{H}$ once
 a set of $\langle\{\sigma_{i}\}\rangle$ is known from either experiment or simulation.
A generalization 
of this approach to several such variables will be described later in this letter. 
Under such assumptions, the Hamiltonian, $\mathcal{H}$ is rewritten as
$\mathcal{H}=\mathcal{H}_{ref}+\sum_{i=1}^{K}{\lambda_{i}\sigma_{i}}$ where 
the expectation values of $K$ order parameters are known, $\mathcal{H}_{ref}$ is a reference 
Hamiltonian and $\lambda$ is a coupling constant. For simplicity in presentation, 
the theory is developed for a single order parameter, $\sigma\equiv\sigma_{i}$ and $K=1$. Then 
the expectation value can be written as: 
\begin{equation}
\label{derivative}
\langle\sigma\rangle =- \ \frac{1}{\beta}\ \frac{\partial}{\partial
  \lambda}\ln \int \mathrm{d} \vec{R}^{N}  \mathrm{d} \vec{P}^{N}\ e^{-\beta
  [\mathcal{H}_{ref}(\vec{R}^{N},\vec{P}^{N})+\lambda \sigma]}{\Bigg \vert}_{\lambda=0}
\end{equation}
The construction of a CG model can be thought of as equivalent to a minimization of 
the difference between the target $\langle\sigma\rangle$ and the value
$\langle\sigma^{CG}\rangle$ computed from the CG Hamiltonian
$\mathcal{H}$, namely:
\begin{eqnarray}
\label{constitutiveEq}
 \langle\sigma^{CG}\rangle-\langle\sigma\rangle &=& -\frac{\partial}{\partial
  \lambda}\Bigg\{\frac{1}{\beta}\ln\int{\mathrm{d}\vec{R}^{N}\mathrm{d}\vec{P}^{N}e^{-\beta
  [\mathcal{H}_{ref}(\vec{R}^{N},\vec{P}^{N})+\lambda(\sigma -\langle\sigma\rangle)]}}\Bigg\} \nonumber\\
& = & 0
\end{eqnarray}
 where the condition $\lambda=0$ was dropped. This equation should be
 regarded as a constitutive equation for CG models. Its meaning is
 straightforward: once $\mathcal{H}_{ref}$ is defined, the aim of
 the method is to find an optimal value for $\lambda$ that can best
approximate the required $\langle\sigma\rangle$. 
Another way to look at the problem is to recast Eq.(\ref{constitutiveEq}) in the following 
manner- If the expression inside the braces in Eq.(\ref{constitutiveEq}) is considered to be 
a free energy function, $F$, then Eq. (\ref{constitutiveEq})  has the form $\frac{\partial F}{\partial \lambda}=0$.

This form of Eq.(\ref{constitutiveEq}) suggests that the construction of a model representation can 
be pictured as follows. The observed order parameter average can be thought of as being induced by 
the perturbation of a reference system given by the Hamiltonian $\mathcal{H}_{ref}$. This external 
perturbation is "conjugate'' to the order parameter function, and finding a solution to Eq.(\ref{constitutiveEq}) 
is equivalent to tuning the external perturbation of the reference system so as to obtain the target order 
parameter ensemble average. Thus, in order to obtain the required response $\sigma$ of the system, 
$\lambda$ is tuned. In the spirit of linear response theory, assuming that the
perturbation is small, it is possible to write a Taylor expansion of
Eq.(\ref{constitutiveEq}) truncated at first order:
\begin{equation}
\label{RMC}
 \langle\sigma\rangle-\langle\sigma\rangle[0] = \lambda_1
 \frac{\partial (\langle\sigma\rangle[\lambda])}{\partial
  \lambda}{\Bigg \vert}_{\lambda=0} + O(\lambda_1^2) 
\end{equation}
where it is assumed that $\lambda=\lambda_{1}$ such that $\langle\sigma\rangle[\lambda_1]\simeq\langle\sigma\rangle$. 
Eq.(\ref{RMC}) can be solved iteratively.  When the order parameter $\sigma$ is chosen to correspond to the pair distribution 
function, $g(r)$ of a system, Eq.(\ref{RMC}) becomes equivalent to the reverse monte carlo method.\cite{Lyub,muller} 
Thus, the reverse monte carlo approach is a natural consequence of the constitutive 
equation, Eq.(\ref{constitutiveEq}), when this equation is used to obtain a microscopic CG 
model that reproduces given structural expectation values as represented by the pair 
distribution functions. 
Further elaboration of this method can be found in the literature.\cite{Lyub,muller} 

%============================ Covariance Matrix

\par Covariance matrices are a particular and important example of such correlation functions, 
and they are of critical importance for dimensionality reduction and data analysis in several fields\cite{pcabk,amadei}. Thus, 
in general, if observations are made regarding the values of some function of the coordinates, 
$\{\phi_i(\vec{r}^{n})\}$,  then correlations $\langle\mathscr{C}[\phi_{i},\phi_{j}]\rangle$ between such 
functions can be written as:
\begin{equation}
\mathscr{C}[\phi_{i},\phi_{j}] \equiv \phi_{i}(\vec{r}^{n})\phi_{j}(\vec{r}^{n})
\end{equation}
Thus, the goal becomes one wherein these expectation values have to be reproduced by 
the coarse grained potential. In the theory presented here, the CG potential energy for this 
system becomes:
\begin{equation}
V_{\mathscr{C}} = 
\sum_{i,j}{v_{ij}(\vec{r}^{n})\mathscr{C}[\phi_{i},\phi_{j}]}
\end{equation}
Thus, Eq.(\ref{constitutiveEq}) requires the construction of potentials
$v_{ij}({\vec{r}}^{n})$ such that they satisfy the 
equation:
\begin{equation}
-\frac{1}{\beta}\frac{\partial}{\partial v_{ij}(\vec{r}^{n})}\ln\Big[\int{\exp\Big\{-\beta V_{\mathscr{C}}\Big\}
d\vec{r}^{n}}\Big] = \langle\mathscr{C}[\phi_{i},\phi_{j}]\rangle
\end{equation}
An approximate solution of this equation can be obtained through the iterative 
reverse monte carlo technique described previously (Eq. \ref{RMC}). An important special 
case involves the construction of CG hamiltonians that reproduce the covariance 
matrix of displacements, $\phi_{i}(\vec{r})=\vec{r}_{i}-\langle\vec{r}_{i}\rangle$, of atoms 
or sites labeled $i$, generated by a molecular ensemble. Such a covariance matrix is 
used to characterize conformational substates, and their collective fluctuation modes in proteins 
and other biomolecules. We now demonstrate that this approach naturally leads to the construction 
of double and multiple well network models\cite{maragakis,best} that have been proposed 
under phenomenological considerations, to describe mechanisms by which proteins undergo large 
scale conformational change. For simplicity, consider a situation wherein a protein structure is 
found to exist in two distinct conformations, which can be distinguished from each other by the value 
of some macroscopic order parameter, $\xi$. For each of these values of $\xi$ a different covariance 
matrix can be constructed. Then, one can construct a potential that reproduces the observed or simulated set 
of position covariances, $\mathscr{C}_{\xi}[\vec{r}_{i},\vec{r}_{j}]$ , or as a special case, root mean 
square fluctuations (B-factors) estimated from experiment, for each intermediate conformation 
(i.e each value of the order parameters, $\xi$). Thus, the coarse-grained potential that reproduces these 
covariances becomes:
\begin{equation}
V[\vec{r}^{n}\mid\xi] = \sum_{i,j}{v_{ij}[\xi]\delta\vec{r}_{i}\cdot\delta\vec{r}_{j}}
\end{equation}
Experimental B-factors of proteins, that are proportional to mean square fluctuations of residues (i.e
diagonal elements of the protein backbone covariance matrix) are often directly measured. Thus, a special 
case of such models is to construct CG Hamiltonians such that they reproduce experimentally measured B-factors 
of protein residues. An additional condition is imposed that the CG model of the protein structures retain the 
same topology as that observed in the original protein conformations \cite{tirion}. With 
these conditions, if only two distinct conformational substates of the protein, characterized by a single order 
parameter $\xi\in [0,1]$ are considered, the "double well" network models \cite{maragakis} used in the literature
are recovered; Namely:
\begin{eqnarray} 
v_{ij}[\xi] & = v_{0}\Gamma_{ij}[0]; & \xi \in [0,0.5]\nonumber\\
              & = v_{1}\Gamma_{ij}[1];  & \xi \geq 0.5   
\end{eqnarray}
Here, $v_{0}$ and $v_{1}$ are constant and $\Gamma_{ij}[\xi]$ is the connectivity matrix of the conformational substate $\xi$, defined as $\Gamma_{ij} = 1$
if amino acid backbone atoms, $i$ and $j$ are within a prescribed cutoff, and zero otherwise.\cite{tirion}  
Thus, this theory provides a rigorous statistical mechanical framework for the 
derivation of coarse-grained models that can be used to describe protein conformational 
transitions. 
%============================ Covariance matrix ==== END

It has been demonstrated that the theory presented can be used to construct microscopic 
statistical mechanical ensembles which reproduce low resolution information in the form of given 
expectation values of a few discrete order parameters. It will now be shown that the same 
theory can be utilized to perform the opposite and important process of reduction, wherein molecular 
coarse grained models are constructed from an equilibrium statistical mechanical ensemble of 
a complex system. This requires the application of the theory to the reproduction of continuous order
parameter expectation values, since such reduction processes generally involve the construction of
reduced many particle distribution functions, rather than discrete expectation values. 
For this purpose, the fundamental equation, Eq.(\ref{constitutiveEq}), is still valid once the following
transformation is applied:
\begin{equation}
\label{translate}
\lambda\sigma \longrightarrow \int \mathrm{d} \vec{\mathcal{R}}^{'N}\mathcal{V}_{\sigma}(\vec{\mathcal{R}}^{'N})\sigma(\vec{\mathcal{R}}^{'N}\vert\vec{\mathcal{R}}^N ) 
\end{equation}
Correspondingly, all variations in the coupling constants, $\lambda$, become functional derivatives:
\begin{equation}
\frac{\partial}{\partial\lambda}\rightarrow \frac{\delta}{\delta\mathcal{V}_{\sigma}(\vec{\mathcal{R}}^{N})}
\end{equation}
The "external potential" $\mathcal{V}_{\sigma}(\vec{\mathcal{R}}^{N})$, Eq. (\ref{translate}), will be used with the reference Hamiltonian, 
$\mathcal{H}_{ref}(\vec{P}^{N}) = \sum_{I=1}^{N} \frac{\vec{P}_i^2}{2M_i}$, where $M_{I}$ 
is the mass associated with the $I$-th particle (site). In Eq.(\ref{translate}) the
symbol $\vec{\mathcal{R}}^N$ is employed instead of $\vec{R}^N$ in order
to differentiate between the treatment of inverse methods and methods for
reducing the degrees of freedom. In the latter the atomistic molecular ensemble
of the system is known, and the target of this methodology is to build
a new coarse grained model associated with a simplified topology of the original 
system, which is consistent with the original molecular ensemble \cite{Voth1}. 
To construct such a model, a mapping of the position of the atomistic coordinates
${\vec{r}}^n$ to CG sites in the simplified topology, $\vec{\mathcal{R}^{N}}$, is defined 
through a mapping operator $M_{\vec{\mathcal{R}}^{N}}({\vec{r}}^n)$. This mapping 
operator relates the two sets of coordinates through
\begin{equation}
\vec{\mathcal{R}}^{N}=M_{\vec{\mathcal{R}}}({\vec{r}}^n).
\end{equation}
An example of such a mapping is one where the coordinates $\vec{r}$ of the 
atoms are mapped onto their center of mass. 
The corresponding CG potential defined by the right side of 
Eq. (\ref{translate}) has a simple physical interpretation. 
The integral computes the potential energy of a particular 
configuration (or ``snapshot'') of the system. Each such configuration 
is represented by the variable $\vec{\mathcal{R}}^N$. 
Since the order parameter, $\sigma(\vec{\mathcal{R}}^{'N}\vert\vec{\mathcal{R}}^{N})$ 
is a continuous function defined over the coarse grained space, the potential 
energy due to the atoms at positions $\vec{\mathcal{R}}^{N}$ is integrated over 
all the spatial degrees of freedom in Eq. (\ref{translate}). The energy of each such snapshot 
can then be written as the sum over interactions between each atom at the position 
$\vec{\mathcal{R}}^{'N}$  and a corresponding external field 
$\mathcal{V}_{\sigma}(\vec{\mathcal{R}}^{'N})$. The external field couples to the atoms 
through the order parameter $\sigma(\vec{\mathcal{R}}^{'N}\vert\vec{\mathcal{R}}^N )$: 
for example, an electric field will interact with the charge of atoms, whereas a 
magnetic field will affect the corresponding spin state of each atom. This potential 
energy associated with a particular configuration (or ``snapshot'') $\vec{\mathcal{R}}^N$ 
is the ``mean-field''. 
\par As with models constructed using inverse methods, the problem of 
building a CG model consists in solving Eq. (\ref{constitutiveEq}). If $Z$ is the partition 
function associated with the CG Hamiltonian
\begin{eqnarray}
\mathcal{H}(\vec{R}^{N},\vec{P}^{N}) & =\mathcal{H}_{ref}(\vec{P}^{N})+
\int{\mathrm{d}\vec{\mathcal{R}}^{'N}\mathcal{V}_{\sigma}(\vec{\mathcal{R}}^{'N})\sigma(\vec{\mathcal{R}}^{'N}\vert\vec{\mathcal{R}}^N)}\nonumber\\
&=\mathcal{H}_{ref}(\vec{P}^{N})+\mathcal{H}_{pot}(\vec{\mathcal{R}}^{N})
\end{eqnarray}
 then Eq.(\ref{constitutiveEq}) can be recast as:
\begin{eqnarray}
-\frac{1}{\beta}\frac{\partial\ln Z[\sigma]}{\partial\mathcal{V}_{\sigma}[\vec{\mathcal{R}}_{0}^{N}]}
& = & \frac{1}{Z^{conf}}\int{d\vec{\mathcal{R}}^{N}\sigma(\vec{\mathcal{R}_0}^{N}\mid\vec{\mathcal{R}}^N)
e^{-\beta\mathcal{H}_{pot}(\vec{\mathcal{R}}^{N})}}\nonumber\\ 
 & = & \langle\sigma(\vec{\mathcal{R}}_{0}^{N})\rangle_{AA} . 
\label{final_eq}
\end{eqnarray}  
Here $Z^{conf}[\sigma] \equiv \int \mathrm{d} \vec{\mathcal{R}}^{N}
e^{-\beta \mathcal{H}(\vec{\mathcal{R}}^{N})}$.
This equation can be solved variationally. A specific
choice for $\langle\sigma (\vec{\mathcal{R}_{0}^{N}}) \rangle$ that has
been used in literature is to minimize the difference between the many particle 
reduced distribution function, $\langle\sigma (\vec{\mathcal{R}}_{0}^{N})\rangle_{AA}=\langle
\delta[\vec{\mathcal{R}}_{0}^{N}-M_{\vec{\mathcal{R}}}({\vec{r}}^n)]\rangle$,
and the many particle distribution function in the simplified topology, 
$\langle \sigma (\vec{\mathcal{R}}_{0}^{N}) \rangle_{CG}=\langle
\delta[\vec{\mathcal{R}}_{0}^{N}-\vec{\mathcal{R}}^{'N}]\rangle$ for  $\forall \vec{\mathcal{R}}_{0}^{N} \in \vec{\mathcal{R}}^{N}$.
For this specific selection the effective potential can be written as:
\begin{equation}
\label{eqB}
\mathcal{V}_{\mathcal{R}_0}(\vec{\mathcal{R}}_{0}^{N})  = -k_{B}T\Bigg\{\ln\Bigg[
\int{\mathrm{d}\vec{r}^{n}
\delta[\vec{\mathcal{R}}_{0}^{N}-M_{\vec{\mathcal{R}}}(\vec{r}^{n})]e^{-\beta u(\vec{r}^{n})}}\Bigg]
+\ln\Big[\frac{Z^{conf}}{z^{conf}}\Big]\Bigg\}
\end{equation}
where $z^{conf} \equiv \int \mathrm{d} \vec{r}^{\,n}
e^{-\beta u(\vec{r}^{n})}$ and $u(\vec{r}^{n})$ is the atomistic
potential energy.
Observing that the potential is defined upto a constant, and that 
$k_{B}T\ln\frac{Z^{conf}}{z^{conf}}=const$, this equation corresponds to the 
fundamental result underlying the MS-CG method \cite{Voth1}. Elaborations of this
method can be found in the cited literature. 

\par Both the inverse methods and the CG formulations derived here as specific implementations 
of Eq. (\ref{constitutiveEq}) involve in practice, a minimization of the difference in the expectation 
values of an order parameter when evaluated over a CG ensemble relative to an ensemble average 
measured either from experiment or simulation. A natural means by which such a minimization 
can be achieved is to perform a linear least squares fit of the unknown (coarse-grained) average
to the target value. Such a fitting procedure offers a simple but powerful numerical implementation
of the constitutive equation, Eq.(\ref{constitutiveEq}) and has been extensively used in the context of 
constructing molecular CG models.\cite{Voth1}

\par Eq.(\ref{eqB}) can also be interpreted in terms of the loss of information relative to the 
original atomistic ensemble, upon performing the reduction procedure. This equation states 
that the Boltzmann factor associated with a specific CG conformation
$\vec{\mathcal{R}}_{0}^{N}$ should be equal to the number of microscopic
conformations that can map to it, divided the total number of
microscopic states.
This interpretation of Eq.(\ref{eqB}) suggests an alternative approach 
to construct CG representations through the minimization of the relative 
entropy between the atomistic and lower resolution CG system. 
Such a formalism has been recently employed \cite{Shell} to
derive CG potentials. Within the relative entropy formalism it
can be shown that the minimal entropic contribution due to this loss of information,
$S_{map}$, is given by:
\begin{equation}
\label{Smap}
S_{map}=-k_{B}\ln\Bigg\{\frac{1}{z^{conf}}\int\mathrm{d}\vec{r}^{n}
\delta[\vec{\mathcal{R}}_{0}^{N}-M_{\vec{\mathcal{R}}}({\vec{r}}^n)]e^{-\beta u(\vec{r}^{n})}\Bigg\}
\end{equation}
Thus, the entropic contribution is (upto a constant) nothing but the logarithm of the CG 
potential energy in Eq.(\ref{eqB}). It should be noted that relative entropy arguments can be used 
to recover the Gibbs-Bogoliubov-Feynman thermodynamic inequality that is of crucial importance 
in equilibrium statistical mechanics.\cite{Shell}

%============================ DFT 
As a final example of the generality of the approach presented here, 
it is easy to see that if the ensemble averaged single particle density of 
a system, $\rho(\vec{r}) = \frac{1}{N}\sum_{i=1}^{n}{\langle\delta (\vec{r}-\vec{r}_{i})\rangle}$
is chosen to be the desired target order parameter average, then
Eq.(\ref{constitutiveEq}) implies that this requires the construction of a single particle potential 
$v(\vec{r})\equiv v[\rho](\vec{r})$ that must generate a Boltzmann ensemble which reproduces 
this average density. The potential derived as a result of this requirement is thus a functional of 
the single particle density, and the model formulation that results corresponds to the well 
known classical density functional theory of classical statistical mechanics.\cite{oxtoby,wu} 
%============================ DFT ==== END

%---------------------------------------
%            CONCLUSION          
%---------------------------------------

In conclusion, it has been shown that many commonly used methods to analyse 
classical statistical ensembles are specific realizations of a generalized mean field 
theory that can be used to model these ensembles. The generalized formalism proposed here
also suggests that equilibrium coarse graining methods that are routinely employed in the study 
of complex molecular systems are closely linked to inverse problems involving these systems,
where low resolution information from experiment or simulation is used to reconstruct 
a microscopic description of the underlying physical system. It has also been shown that 
both atomistic density functional theories and lower resolution mean field coarse 
grained models can be derived from this formalism through an appropriate choice of the 
target properties of interest. Thus, the generalized mean field theory discussed in 
this work offers a simple and general framework for both the construction of sophisticated
coarse grained approaches to modeling the statistical mechanical properties of complex 
molecular systems, as well as the reconstruction of microscopic models based on macroscopic 
experimental observations of their ensemble averaged properties.   
\par We thank Hans Andersen and Gregory Voth for valuable comments and discussions. 
%---------------------------------------------
%     Bibliography
%---------------------------------------------

\end{document}